\documentclass[prb,showpacs,twocolumn,aps,superscriptaddress,a4paper,floatfix]{revtex4-1}

\usepackage{color}
\usepackage{dcolumn}
\usepackage{amsmath}
\usepackage{amssymb}
\usepackage{mathtools}
\usepackage{graphicx}
\usepackage{latexsym}
\usepackage{amsfonts}
\usepackage{amssymb}
\usepackage{comment}
\usepackage{natbib}
\DeclareGraphicsExtensions{.pdf,.gif,.jpg}
\usepackage[utf8]{inputenc}
\newcommand{\be}{\begin{equation}}  
\newcommand{\ee}{\end{equation}}
\newcommand{\beq}{\begin{eqnarray}}  
\newcommand{\eeq}{\end{eqnarray}}

\renewcommand\Im{\operatorname{Im}}

\begin{document}


\title{Numerically exact counting statistics of energy current in the Kondo regime}

\author{Michael Ridley}

\affiliation{The Raymond and Beverley Sackler Center for Computational Molecular
and Materials Science, Tel Aviv University, Tel Aviv 6997801, Israel}

\affiliation{School of Chemistry, Tel Aviv University, Tel Aviv 69978, Israel}

\author{Michael Galperin}

\affiliation{Department of Chemistry \& Biochemistry, University of California San Diego, La Jolla, CA 92093, USA}

\author{Emanuel Gull}

\affiliation{Department of Physics, University of Michigan, Ann Arbor, Michigan
48109, USA}
\affiliation{Center for Computational Quantum Physics, Flatiron Institute, New York, New York 10010, USA}

\author{Guy Cohen}

\affiliation{The Raymond and Beverley Sackler Center for Computational Molecular
and Materials Science, Tel Aviv University, Tel Aviv 6997801, Israel}

\affiliation{School of Chemistry, Tel Aviv University, Tel Aviv 69978, Israel}

\date{\today}  

\begin{abstract}
We use the inchworm Quantum Monte Carlo method to investigate the full counting statistics of particle and energy currents in a strongly correlated quantum dot. Our method is used to extract the heat fluctuations and entropy production of a quantum thermoelectric device, as well as cumulants of the particle and energy currents. The energy--particle current cross correlations reveal information on the preparation of the system and the interplay of thermal and electric currents. We furthermore demonstrate the signature of a crossover from Coulomb blockade to Kondo physics in the energy current fluctuations, and show how the conventional master equation approach to full counting statistics systematically fails to capture this crossover.
\end{abstract}

  
\maketitle  


\section{Introduction}

Energy transport through small junctions is a major paradigm at the heart of attempts to formulate thermodynamic principles applicable to nanoscale quantum systems.~\cite{allahverdyan_breakdown_2001,martinez_dynamics_2013,ludovico_dynamical_2014,esposito_nature_2015,bruch_quantum_2016,ochoa_energy_2016,haughian_quantum_2018}
The latter are essential for understanding and improving operational efficiency in nanoelectronic devices,\cite{pop_energy_2010} where quantum entanglement effects are inseparably intertwined with performance fluctuations.~\cite{kosloff_quantum_2014,uzdin_equivalence_2015,esposito_efficiency_2015,niedenzu_quantum_2018}

\begin{figure}
\includegraphics{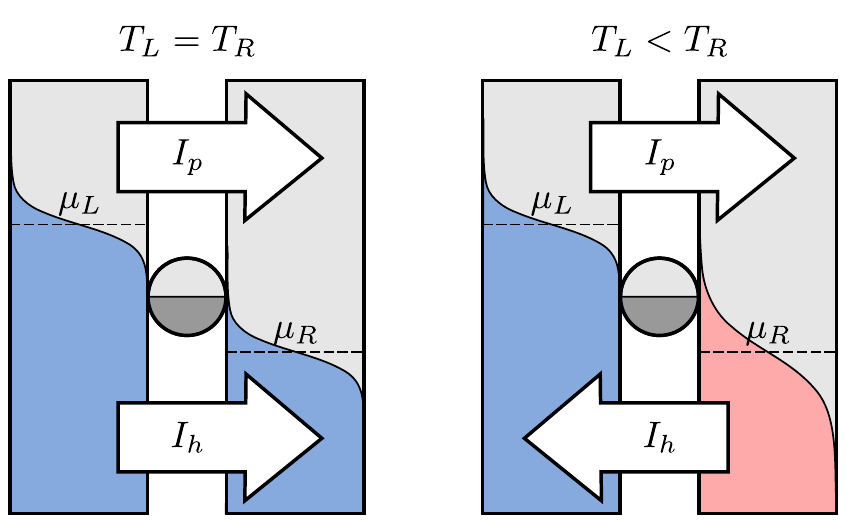}
\caption{Illustration of a quantum junction comprising leads $L$ and $R$ coupled by a central quantum dot region $D$, with a chemical potential bias at no temperature bias (left) and an opposite temperature bias (right). Depending on the choice of parameters, the particle and heat currents, $I^{p}$ and $I^{h}$, may be expected to flow in either the same direction (left) or opposite directions (right).}
\label{fig:nanojunction} 
\end{figure}

Fluctuations in the energy and particle currents~\cite{blanter_shot_2000, pietzonka_universal_2018} can be studied through consideration of the full counting statistics (FCS) of particle and energy transfer.~\cite{levitov_electron_1996,nazarov_full_2007,esposito_entropy_2007,bednorz_formulation_2008}
A particularly challenging problem in this field is the study of electronic correlations and their impact on the performance of quantum devices coupled 
to conducting reservoirs.
A schematic view of a junction is shown in Fig.~\ref{fig:nanojunction}.
This comprises a central molecule or quantum dot where confined electrons interact strongly, contacted by weakly interacting metallic left ($L$) and right ($R$) reservoirs.
A voltage bias is applied across the junction, and an additional thermal or temperature bias may either not be applied (left panel) or applied (right panel).

Under certain conditions, the system may be characterized by a threshold known as the Kondo temperature $T_{K}$.
Below this threshold, transport properties are dominated by the formation of a correlated resonance in the conductance spectrum.\cite{haldane_scaling_1978,hewson_kondo_1993}
A bias voltage (or chemical potential difference between the leads) splits the Kondo resonance into a pair of peaks 
centered near the lead chemical potentials.~\cite{meir_low-temperature_1993,cronenwett_tunable_1998,schmidt_transport_2010}
The resonance width is set by $T_{K}$, and it is typically much sharper than features related to noninteracting resonant tunneling,~\cite{ng_-site_1988} as seen in experiments on strongly correlated single-molecule transistors.~\cite{cronenwett_tunable_1998,liang_kondo_2002}

A thermal bias (right panel of Fig.~\ref{fig:nanojunction}) may be applied to the junction by keeping the two reservoirs at different temperatures $T_{L}$ and $T_{R}$. The interplay between electric and thermal biases can be used to implement \textit{e.g.} heat engines, heat pumps and refrigerators, depending on the direction of the thermal gradient and the relative direction of flow of particles $I_{p}$ and heat $I_{h}$.~\cite{kosloff_quantum_2014,whitney_finding_2015, benenti_fundamental_2017, whitney_quantum_2018} 

Recent experiments on molecular junctions show that both the Seebeck and Peltier coefficients can be measured at the nanoscale~\cite{reddy_thermoelectricity_2007,lee_heat_2013,brechet_evidence_2013,kim_electrostatic_2014,rincon-garcia_thermopower_2016,cui_perspective:_2017,cui_peltier_2018}
and the quantized character of thermal transport in nanojunctions was demonstrated.~\cite{schwab_measurement_2000,cui_quantized_2017} In addition, a recent measurement of thermal gradient induced 
particle
noise was reported in the literature.~\cite{lumbroso_electronic_2018} 


In the linear regime, quantum coherence universally weakens heat engine performance,~\cite{brandner_universal_2017} and reductions in power and efficiency have also been reported in nonlinear quantum heat engines.~\cite{karimi_otto_2016}
However, there is no comprehensive answer to the question of whether quantum effects improve or reduce the performance of nonlinear heat engines at the nanoscale.
In certain specific cases, quantum coherence effects have been shown to increase the thermal efficiency,~\cite{chen_molecular_2017} to exceed the Carnot efficiency and approach a perfect efficiency of 1,~\cite{toyabe_nonequilibrium_2010} to increase the power output \cite{scully_quantum_2011} and to reduce fluctuations in the power. \cite{ptaszynski_coherence-enhanced_2018} 


Whereas it is possible to study nanoscale thermoelectricity within a single-particle approximation with Green's function \cite{crepieux_enhanced_2011,ludovico_dynamics_2016,covito_transient_2018} or DFT-based \cite{pauly_length-dependent_2008,zotti_heat_2014,eich_density-functional_2014} methods, the effect of Kondo physics on the thermoelectric performance of devices is not fully understood beyond linear response. An exception is recent work based on the matrix product states method, which gives access to the current generated by a temperature bias in the Kondo regime.~\cite{dorda_thermoelectric_2016} In related work, simulations of the current within the non-crossing approximation show a strong enhancement in the Seebeck coefficient by asymmetric dot--lead couplings in the Kondo regime.~\cite{perez_daroca_enhancing_2018}

Going beyond the study of heat and charge currents, one may investigate higher-order correlations in the transferred charge and energy. These higher order statistical cumulants, and their underlying probability distributions, are obtained from the cumulant generating function for the FCS, and can reveal a whole range of information not present in the first moment.~\cite{landauer_noise_1998} This includes particle traversal times,~\cite{ridley_partition-free_2017} waiting time distributions \cite{tang_waiting_2014,rudge_distribution_2016,ptaszynski_waiting_2017,ridley_numerically_2018,kosov_waiting_2018} shot noise \cite{buttiker_scattering_1990,levitov_electron_1996,blanter_shot_2000} and associated quasiparticle charges.~\cite{yamauchi_evolution_2011} In addition, the steady state cumulant generating function can be related to fluctuations in the thermal efficiency.~\cite{esposito_efficiency_2015} 

To date, the majority of the exact FCS theory has focused on noninteracting models of nano junctions, beginning with the work of Levitov and Lesovik on the counting statistics of charge transport in the steady state regime, which reduces to the Landauer--B{\"u}ttiker theory of shot noise, in which the propagating charge carriers are assumed to be coherent waves in the device region.~\cite{landauer_noise_1998,blanter_shot_2000,ouyang_quantum_2018} This program culminated with the path integral nonequilibrium Green's function (PINEGF) approach pioneered by Tang \textit{et al.}, ~\cite{tang_full-counting_2014,tang_waiting_2014,yu_full-counting_2016,tang_full-counting_2017,tang_thermodynamics_2018} valid for FCS calculations in the transient regime following a quench and for arbitrary time-dependent driving fields. 

In strongly correlated systems, the quantum master equation (QME) approach has proven popular for the investigation of charge~\cite{flindt_universal_2009,flindt_counting_2010,albert_electron_2012} and energy~\cite{esposito_entropy_2007,clerk_full_2011,sanchez_detection_2012,silaev_lindblad-equation_2014,ptaszynski_coherence-enhanced_2018} FCS in Coulomb-blockaded systems. It lends itself well to analytical investigation, so it has been used to derive exact relations for optimized thermal efficiencies in molecular heat engines~\cite{esposito_thermoelectric_2009, abah_efficiency_2014} and exact fluctuation relations in the counting statistics of heat exchange.~\cite{friedman_quantum_2018} However, the QME method is generally limited to the weak impurity--bath coupling regime: $\Gamma$, the typical strength of hybridization between the lead states and those of the molecule or dot, must be substantially smaller than all other energy scales in the problem. In recent years, the QME method was extended beyond the sequential tunneling Coulomb-blockade regime to include spin-flip scattering cotunneling processes, so that noise and FCS can be studied at higher values of $\Gamma$.~\cite{thielmann_cotunneling_2005,leijnse_kinetic_2008,kaasbjerg_full_2015,seja_violation_2016,walldorf_thermoelectrics_2017} 

However, the Kondo regime and current noise resulting from the formation of a Kondo peak ~\cite{hewson_kondo_1993,cohen_greens_2014} are inaccessible to perturbative QME approaches.~\cite{miwa_towards_2017}
Indeed, the study of current fluctuations in the Kondo regime has mostly been carried out by methods specialized to very specific parameter regimes.~\cite{sela_fractional_2006,gogolin_towards_2006,carr_full_2011} The situation has changed with the recent development of a numerically exact method for the computation of full particle counting statistics (FCS).~\cite{ridley_numerically_2018} In this context, `numerically exact' refers to the controlled evaluation of a quantity with known error bounds that can be made as small as desired with enough computer time. This approach was based on the inchworm quantum Monte Carlo (iQMC) algorithm,~\cite{cohen_taming_2015,antipov_currents_2017,chen_inchworm_2017-1,chen_inchworm_2017} a powerful method for the stochastic evaluation of propagators in strongly correlated impurity models. It circumvents the dynamical sign problem that plagues conventional QMC methods\cite{gull_numerically_2011} and scales polynomially in time.~\cite{cohen_taming_2015} In Ref.~\onlinecite{ridley_numerically_2018}, numerically exact calculations of the current noise were presented, showing a crossover from interaction-induced suppression to enhancement as the temperature was lowered below $T_{K}$. This finding corroborates the experimental data showing Kondo-enhanced noise in strongly interacting impurity models, ~\cite{delattre_noisy_2009} and also indicates the failure of noninteracting theory to properly account for noise in the Kondo regime.

In the present study we extend the iQMC FCS method to the full counting statistics of nonequilibrium energy transfer. We treat junctions with electron--electron interactions, thus taking steps towards a sorely-needed quantum theory of thermoelectricity.~\cite{whitney_quantum_2018}
Our methodology enables numerically exact calculations of cumulants of charge, energy and heat transfer in strongly correlated regimes, as well as cross-correlations between particle and energy transfer. Through these quantities, we obtain access to the entropy production.

The structure of the paper is as follows: in Section~\ref{model} we introduce the model of the quantum junction utilized in this study.
Section~\ref{FCS} introduces the techniques we use (both numerically exact iQMC and approximate QME) to calculate particle and energy transfer statistics in the transient regime, and Section~\ref{cumulants} applies this to quantities of interest in the study of thermoelectric properties of junctions.
In Section~\ref{results} we present our main results.
This includes a comparison with QME data, a discussion of the effects of interaction and temperature on heat cumulants and associated heat current fluctuations, and calculations of entropy production.
Finally, our conclusions can be found in Section~\ref{conclusion}.


\section{The Model}\label{model}
We consider a junction described by the nonequilibrium Anderson impurity model (AIM) Hamiltonian,
\begin{equation}
\hat{H}=\hat{H}_{D}+\hat{H}_{B}+\hat{H}_{V}.
\label{eq:Anderson}
\end{equation}
Here, $\hat{H}_{D}$ describes a dot (or molecule) region with on-site interactions;  $\hat{H}_{B}=\sum_\ell{\hat{H}_{\ell}}$ describes two large noninteracting reservoirs or leads, denoted by $\ell\in\{L,R\}$; and $\hat{H}_{V}$ is a hybridization Hamiltonian coupling the dot to both leads. 
These terms are given by
\begin{align}
\hat{H}_{D} & =\underset{\sigma}{\sum}\varepsilon_{\sigma}\hat{d}_{\sigma}^{\dagger}\hat{d}_{\sigma}+U\hat{d}_{\uparrow}^{\dagger}\hat{d}_{\uparrow}\hat{d}_{\downarrow}^{\dagger}\hat{d}_{\downarrow},\label{eq:Hdot}\\
\hat{H}_{B} & =\underset{k\sigma,\ell}{\sum}\varepsilon_{k\sigma,\ell}\hat{a}_{k\sigma,\ell}^{\dagger}\hat{a}_{k\sigma,\ell},\label{eq:Hbath}\\
\hat{H}_{V} & =\underset{k\sigma,\ell}{\sum}\left(V_{k\sigma,\ell}\hat{a}_{k\sigma,\ell}^{\dagger}\hat{d}_{\sigma}+\textrm{H.c.}\right).\label{eq:Hhyb}
\end{align}
The $\hat{d}^\dagger_{\sigma}$ ($\hat{d}_{\sigma}$) operators create (annihilate) an electron with spin $\sigma\in\left\{\ \uparrow, \downarrow \right\}$ on the dot, and the $\hat{a}_{k\sigma,\ell}^\dagger$ ($\hat{a}_{k\sigma,\ell}$) perform the analogous operation on the single particle level $k$ of lead $\ell$. $\varepsilon_{\sigma}$ and $\varepsilon_{k\sigma,\ell}$ are single particle energies on the dot and lead levels, respectively, and $U$ sets the strength of Coulomb repulsion on the dot. Throughout this work, we impose particle--hole symmetry by setting $\epsilon_{\sigma}=-\frac{U}{2}$. Finally, the $V_{k\sigma}$ control tunneling between the dot and leads.

We will simulate dynamics starting from an initial state $\rho_{0}=\rho_{L}\otimes\rho_{D}\otimes\rho_{R}$, where each lead $\ell$ is in a thermal state characterized by a temperature $T_\ell=1/\beta_\ell$ and a chemical potential $\mu_{L/R}=\pm V/2$.
We assume that the dot is prepared in one of the four eigenstates of $\hat{H}_{D}$: the empty state $\left|0\right\rangle$, full state $\left|\uparrow\downarrow\right\rangle$, or one of the two magnetized half-occupied states $\left|\sigma\right\rangle$.
In what follows these local or `atomic' states are referred to collectively by the index $\phi$. At time $t=0$, the system begins to evolve with respect to $\hat{H}$, corresponding to instantaneously adding the hybridization term $\hat{H}_{V}$ to the Hamiltonian to model a sudden coupling of the leads to the dot. This setup is often referred to as either a coupling quench or a partitioned approach to the switch-on problem.~\cite{ridley_formal_2018}

The $\varepsilon_{k\sigma,\ell}$ and $V_{k\sigma,\ell}$ are effectively set by the coupling density $\Gamma_{\ell}\left(\omega\right)$ that describes the electron escape rate of lead $\ell$:
\begin{equation}
\Gamma_{\ell}\left(\omega\right)=\pi\underset{k\sigma}{\sum}\left|V_{k\sigma,\ell}\right|^{2}\delta\left(\omega-\varepsilon_{k\sigma,\ell}\right).\label{eq:density}
\end{equation}
We take this to be a flat band with a smooth cutoff,\cite{werner_diagrammatic_2009}
\begin{equation}
\Gamma_{\ell}\left(\omega\right)=\frac{\Gamma_{\ell}}{\left(1+e^{\nu\left(\omega-\Omega_{c}\right)}\right)\left(1+e^{-\nu\left(\omega+\Omega_{c}\right)}\right)}.\label{eq:fermdensity}
\end{equation}
In what follows, we set $\Gamma_{\ell}=\frac{1}{2}$ such that $\Gamma\equiv\underset{\ell}{\sum}\Gamma_{\ell}=1$ determines our unit of energy. We take the leads' band cutoff $\Omega_{c}$ to be $10\Gamma$, and their edge width $\frac{1}{\nu}$ to be $\Gamma/10$.

\section{FCS of energy and particles}\label{FCS}

A chemical potential bias and/or a temperature difference between the leads engenders charge and energy fluxes across the junction.
For the interface between any lead $\ell$ and the dot, FCS yields a way to study the statistics of both electron transfer, $\Delta\hat N_\ell\left(t\right)\equiv \hat N_\ell\left(t\right)-\hat N_\ell\left(0\right)$;
and energy transfer, $\Delta \hat H_\ell\left(t\right)\equiv \hat H_\ell\left(t\right)-\hat H_\ell\left(0\right)$.
Here $\hat N_\ell\left(t\right)$ and $\hat H_\ell\left(t\right)$, respectively, are operators corresponding to the total number of electrons and total energy in lead $\ell$.
In the following section, we briefly introduce some of the main concepts and definitions of FCS for the convenience of the reader.

\subsection{FCS from iQMC}\label{FCS_iQMC}

Within the FCS formalism, the statistics of currents at the interface with lead $\ell$ are obtained from a two-point measurement of the total number of particles or energy within reservoir $\ell$ at times $0$ and $t$.
Multiple realizations of the quantum process yield information on the probability $P\left(\Delta n_\ell,\Delta\epsilon_\ell,t\right)$ to observe transfer of $\Delta n_\ell$ electrons and $\Delta \epsilon_\ell$ energy across the interface.
These probabilities are encoded within the generating function~\cite{esposito_nonequilibrium_2009, tang_full-counting_2014}
\begin{equation}
Z\left(t;\lambda,\chi\right)\equiv\underset{\triangle n_{\ell},\triangle\varepsilon_{\ell}}{\sum}P\left(\triangle n_{\ell},\triangle\varepsilon_{\ell},t\right)e^{i\lambda\triangle n_{\ell}}e^{i\chi\triangle\varepsilon_{\ell}}.
\label{eq:Z_def-1}
\end{equation}
The Fourier variables $\lambda$ and $\chi$ are called counting fields.
In general there may be a separate counting field for every lead in a multiterminal counting experiment; the generalization of Eq.~(\ref{eq:Z_def-1}) to the multiterminal case is straightforward.

Assuming the initial state of the system to be an eigenstate of the decoupled dot Hamiltonian, Eq.~\eqref{eq:Z_def-1} can be expressed in terms of counting-field-modified propagators
\begin{equation}
\begin{aligned}
Z\left(t;\lambda,\chi\right) &= \textrm{Tr}\left[
\hat{U}_{\left(\lambda,\chi\right)}\left(t,t_0\right)\,\hat{\rho}_{0}\,\hat{U}_{\left(\lambda,\chi\right)}^{\dagger}\left(t,t_0\right)
\right]
\\
&\equiv  \textrm{Tr}\left[
T_\mathcal{C}\, e^{-i\int_\mathcal{C}dz\, \hat H_{\left(\lambda,\chi\right)}(z)}\hat \rho_0
\right],
\label{eq:Z_def-2}
\end{aligned}
\end{equation}
where $\mathcal{C}$ is the Keldysh contour comprising a forward branch $\mathcal{C}_{+}$ (from $t_0$ to $t$) followed by a backward branch  $\mathcal{C}_{-}$ (from $t$ to $t_0$), $z$ is a contour-time variable, and $T_\mathcal{C}$ is the time ordering operator on the contour.
Time evolution is governed by the modified propagator and Hamiltonian, respectively given by
\begin{equation}
\hat{U}_{\left(\lambda,\chi\right)}\left(t,t_{0}\right)=e^{i\frac{\chi}{2}\hat{H}_{\ell}}e^{i\frac{\lambda}{2}\hat{N}_{\ell}}\hat{U}\left(t,t_{0}\right)e^{-i\frac{\lambda}{2}\hat{N}_{\ell}}e^{-i\frac{\chi}{2}\hat{H}_{\ell}}
\end{equation}
and
\begin{equation}
\hat{H}_{\left(\lambda,\chi\right)}\left(z\right)=\hat{H}_{D}\left(z\right)+\hat{H}_{B}\left(z\right)+\hat{H}_{V\,\left(s_{\mathcal{C}}\lambda,s_{\mathcal{C}}\chi\right)}\left(z\right).
\end{equation}
Here, $s_\mathcal{C}=\pm 1$ at the $\mathcal{C}_\pm$ branch of the contour and
\begin{equation}
\hat{H}_{V\, \left(\lambda,\chi\right)} = 
\sum_{k\sigma,\ell}\left(V_{k\sigma,\ell}\hat{a}_{k\sigma,\ell}^{\dagger}\hat{d}_{\sigma}
e^{(\lambda+\varepsilon_{k\sigma,\ell}\chi)/2} +\textrm{H.c.}
\right).
\label{Vdressed}
\end{equation}

The object evaluated in the iQMC method is the auxiliary-field-modified propagator between two points $z_{1}$ and $z_{2}$ on the contour,
\begin{equation}
\begin{aligned}p\left(z_{1},z_{2};\lambda,\chi\right)\equiv & \textrm{Tr}_{B}\Bigg[\rho_{0}T_{\mathcal{C}}e^{-i\int_{z_{1}}^{z_{2}}dz\,\hat{H}_{\left(\lambda,\chi\right)}\left(z\right)}\Biggr],\\
p_{\phi\phi'}\left(z_{1},z_{2};\lambda,\chi\right)\equiv & \left\langle \phi\right|p\left(z_{1},z_{2};\lambda,\chi\right)\left|\phi'\right\rangle ,
\end{aligned}
\label{eq:atomicpropagator}
\end{equation}
where the second line describes projection to the atomic $\phi$ basis.

The propagator of Eq.~\eqref{eq:atomicpropagator} is analogous to propagators used in other real time, continuous time hybridization-expansion Monte Carlo methods: whereas earlier implementations only implicitly relied on such propagators,\cite{muhlbacher_real-time_2008,werner_diagrammatic_2009,schiro_real-time_2009,schiro_real-time_2010,cohen_memory_2011} later work explicitly took advantage of their properties.\cite{cohen_taming_2015,antipov_currents_2017,chen_inchworm_2017-1,chen_inchworm_2017,dong_quantum_2017,boag_inclusion-exclusion_2018,krivenko_dynamics_2019,gull_numerically_2011,cohen_numerically_2013,cohen_greens_2014,cohen_greens_2014-2, kubiczek_exact_2019} Here, the hybridization Eq.~\eqref{Vdressed} is modified with respect to its physical counterpart $\hat{H}_{V\, \left(\lambda=0,\chi=0\right)}$ by $\lambda$- and $\chi$-dependent factors.
The first of these modifications was already present in the previous iQMC FCS paper,\cite{ridley_numerically_2018} where it was used to address particle (but not energy) counting statistics.

The bare Monte Carlo process associated with evaluating Eq.~\eqref{eq:atomicpropagator} in the hybridization expansion results in a dynamical sign problem, such that the computational cost of the simulation increases exponentially with time.
The inchworm algorithm\cite{cohen_taming_2015} overcomes this for at least some parameters, such that the FCS can be obtained efficiently.\cite{ridley_numerically_2018}
In the present work, no new conceptual difficulties beyond the introduction of a second counting field (see below) arise.
We therefore refer the reader to the literature for a thorough discussion of how iQMC can be applied to particle counting statistics.\cite{ridley_numerically_2018}
In the rest of this chapter, we limit ourselves to a self-contained description of the the modifications to the theory underlying iQMC, that are needed in order to also access energy counting statistics.

The cumulant generating function at time $t$ can be extracted directly from the propagator between the contour times $t^+$ and $t^-$ corresponding to the physical time $t$ on the two branches of the Keldysh contour:
\begin{equation}
Z\left(t;\lambda,\chi\right)=\sum_\phi p_{\phi\phi}\left(t^{+},t^{-};\lambda,\chi\right).
\label{eq:Zpropagator}
\end{equation}
By expanding Eq.~\eqref{eq:atomicpropagator} in powers of $\hat{H}_{V}$, 
the propagator can be represented as a sum over configurations suitable 
for stochastic Monte Carlo sampling,
\begin{equation}
p_{\phi\phi^\prime}\left(z_1,z_2;\lambda,\chi\right)=\sum_{n=0}^{\infty
}\sum_{\mathcal{C}}w_{\mathrm{loc}}^{\left(n\right)}\left(\mathcal{C}\right)
w_{\mathrm{hyb}}^{\left(n\right)}\left(\mathcal{C};\lambda,\chi\right).
\label{eq:propMC}
\end{equation}
Since only even orders contribute in this combination of expansion, model and propagator, we use $n$ to denote the expansion order, but a term of order $n$ actually contains $2n$ hybridization vertices of the form in Eq.~\eqref{Vdressed}.
These vertices occur at times $\mathcal{C}=\{\zeta_1, \ldots, \zeta_{2n}\}$. The vertex times, all of which are located within the part of the contour between $z_1$ and $z_2$, are integrated over in the second summation.

The $w_{\mathrm{loc}}^{(n)}$ are products of interacting (but purely local) atomic state propagators $p_{\ell\kappa}^{\left(0\right)}\left(z_{1},z_{2}\right)$, defined as in Eq.~\eqref{eq:atomicpropagator} but with $\hat{H}_{\left(\lambda,\chi\right)}$ replaced by $\hat H_0\equiv\hat{H}_{D}+\hat H_{B}$ in the time-ordered integral.
The states $\phi_i$ between each pair of vertices are fully determined by $\mathcal{C}$ and the initial condition $\phi_0$.
Setting $\phi_{2n}=\phi_0$ allows us to write these objects in the following form, which is independent of both $\lambda$ and $\chi$:
\begin{equation}
w_{\mathrm{loc}}^{\left(n\right)}\left(\mathcal{C}\right)=\left(-i\right)^{n_{+}-n_{-}}\prod_{i=0}^{2n-1}p_{\phi_{i}\phi_{i+1}}^{\left(0\right)}\left(z_{i},z_{i+1}\right).
\end{equation}

The $w_{\mathrm{hyb}}^{(n)}$, on the other hand, explicitly depend on the counting fields:
\begin{equation}
w_{\mathrm{hyb}}^{\left(n\right)}\left(\mathcal{C};\lambda,\chi\right)=\sum_{\{\mathbf{X}\}}\mathrm{sign}\left(\mathbf{X}\right)\prod_{i=0}^{n}\tilde{\Delta}^{\lambda,\chi}\left(z_{i},z_{X_{i}}\right).
\end{equation}
In general, $\mathbf{X}$ are all permutations of the integers $(1,\ldots,n)$, and $\tilde{\Delta}^{\lambda,\chi}=\sum_{\ell}\tilde{\Delta}^{\lambda_\ell,\chi_\ell}_{\ell}$, but below we always set $\lambda_\ell=\lambda\delta_{\ell L}$ and $\chi_\ell=\chi\delta_{\ell L}$.
The contribution to the counting-field-modified hybridization from lead $\ell$ is given by
\begin{equation}
\begin{aligned}\tilde{\triangle}_{\ell}^{\lambda,\chi}\left(z_{1},z_{2}\right) & =e^{i\lambda_{\ell}\left(1-\delta_{\nu\nu^{\prime}}\right)}\theta\left(z_{1}-z_{2}\right)\triangle_{\ell}^{>}\left(z_{1}-z_{2};\chi\right)\\
 & +e^{-i\lambda_{\ell}\left(1-\delta_{\nu\nu^{\prime}}\right)}\theta\left(z_{2}-z_{1}\right)\triangle_{\ell}^{<}\left(z_{1}-z_{2};\chi\right),
\end{aligned}
\label{eq:hybKeldysh-1}
\end{equation}
where $\theta$ is the Heaviside function on the contour and $\nu,\nu^{\prime}\in\left\{ \mathcal{C}_{+},\mathcal{C}_{-}\right\}$ are the branch indices of $z_{1}$ and $z_{2}$, respectively.
Finally, the $\lambda$-unmodified lesser and greater hybridization components for lead $\ell$ are obtained from the level width function of Eq.~(\ref{eq:density}):
\begin{equation}
\begin{aligned}\Delta_{\ell}^{>}\left(t_{1}-t_{2};\chi\right) & =i\int_{-\infty}^{\infty}\frac{\mathrm{d}\omega}{\pi}e^{-i\omega\left(t_{1}-t_{2}\right)}e^{i\omega\chi_{\ell}}\\
 & \times\Gamma_{\ell}\left(\omega\right)\left[1-f\left(\omega-\mu_{\ell}\right)\right],\\
\Delta_{\ell}^{<}\left(t_{1}-t_{2};\chi\right) & =-i\int_{-\infty}^{\infty}\frac{\mathrm{d}\omega}{\pi}e^{-i\omega\left(t_{1}-t_{2}\right)}e^{-i\omega\chi_{\ell}}\\
 & \times\Gamma_{\ell}\left(\omega\right)f\left(\omega-\mu_{\ell}\right).
\end{aligned}
\end{equation}
Here $t_{1}$ and $t_{2}$ are physical times corresponding to the contour times $z_{1}$ and $z_{2}$. For leads initially in equilibrium or steady state, the two-time hybridization functions depend only on the time difference $t_{1}-t_{2}$, and it is convenient to use the Fourier shift property 
\begin{equation}
\Delta_{\ell}^{\lessgtr}\left(t_{1}-t_{2};\chi\right) = \Delta_{\ell}^{\lessgtr}\left(t_{1}-t_{2}\mp\chi_\ell\left(1-\delta_{\nu\nu^{\prime}}\right);0\right)
\end{equation}
to avoid numerical issues with strongly oscillating frequency integrals at finite values of $\chi$.

\subsection{FCS from QME}\label{FCS_QME}

The quantum master equation approach to FCS provides an approximate expression for the dynamics of the reduced density matrix modified by the counting field, $\hat{\sigma}\left(t;\lambda,\chi\right) \equiv \mathrm{Tr}_{B} \left\{\hat{\rho}\left(t;\lambda,\chi\right)\right\}$.~\cite{bagrets_full_2003} 

Since the dynamics of the off-diagonal elements of $\hat{\sigma}$ is completely decoupled from that of the diagonal ones in the AIM, 
it is sufficient to consider the diagonal elements of $\hat{\sigma}\left(t;\lambda,\chi\right)$, the counting field-modified ``populations'' $p_{\phi}\left(t;\lambda,\chi\right)$. The generating function $Z\left(t;\lambda,\chi\right)$ is then given by the trace with respect to the dot subsystem~\cite{esposito_nonequilibrium_2009}
\begin{equation}
Z\left(t;\lambda,\chi\right)=\mathrm{Tr}_{S}\left[\hat{\sigma}\left(t;\lambda,\chi\right)\right]=\underset{\phi}{\sum}p_{\phi}\left(t;\lambda,\chi\right).
\label{eq:CGF_QME}
\end{equation}

We collect the $p_{\phi}$ into a vector $\mathbf{p}$, such that---within the QME approximation---$\mathbf{p}$ satisfies the following rate equation:
\begin{equation}
\frac{\partial\mathbf{p}\left(t;\lambda,\chi\right)}{\partial t}=M\left(\lambda , \chi\right) \mathbf{p}\left(t;\lambda,\chi\right).
\label{eq:QME}
\end{equation}
Here, $M$ is a matrix with elements given at $\lambda,\chi=0$ by~\cite{datta_quantum_2005,nitzan_chemical_2013,levy_absence_2019}
\begin{equation}
M_{\phi\phi^{\prime}}=\underset{\ell}{\sum}M_{\phi\phi^{\prime}}^{\ell},
\label{eq:QME_zerolambda_elements}
\end{equation}
where
\begin{equation}
M_{\phi\phi^{\prime}}^{\ell}=\left\{ \begin{array}{cc}
\underset{\phi^{\prime\prime}}{\sum}\Gamma_{\ell}f_{\ell}\left(\Delta E_{\phi^{\prime}\phi^{\prime\prime}}\right) & \phi=\phi^{\prime},\\
-\Gamma_{\ell}f_{\ell}\left(\Delta E_{\phi^{\prime}\phi}\right) & n_{\phi^{\prime}}-n_{\phi}=1,\\
-\Gamma_{\ell}(1-f_{\ell}\left(\Delta E_{\phi\phi^{\prime}}\right)) & n_{\phi^{\prime}}-n_{\phi}=-1,\\
0 & n_{\phi^{\prime}}-n_{\phi}=\pm2,0.
\end{array}\right.
\end{equation}
The terms $M_{\phi\phi^\prime}^{\ell}$ in this expression correspond to transition rates from dot state $\phi^\prime$ to state $\phi$ as mediated by the lead $\ell$, and we have defined $n_\phi$ to be the occupation of the 
decoupled
dot state $\phi$, $\Delta E_{\phi^\prime \phi} \equiv E_{\phi^\prime}-E_{\phi}$ and $f_{\ell}(\epsilon)=1/(1+e^{\beta_\ell(\epsilon-\mu_\ell)})$. The level width $\Gamma_{\ell}$ is taken to be frequency independent, an additional approximation which could be easily removed, but is justified for the band width and shape used here.~\cite{covito_transient_2018,ridley_lead_2019}

In the presence of the counting field, the transition rates undergo the modification
\begin{equation}
M_{\phi\phi^{\prime}}^{\ell}\rightarrow M_{\phi\phi^{\prime}}^{\ell}e^{i\lambda\epsilon_{\phi\phi^{\prime}}}e^{-i\chi\Delta E_{\phi^{\prime}\phi}}.
\label{eq:modified_gamma}
\end{equation}
Here, $\epsilon_{ij}=\pm1$ for $n_{i}=n_{j}\pm1$, and $\epsilon_{ij}=0$ otherwise.
The sign of the exponential factor $-i\chi\Delta E_{\phi^\prime\phi}$ arises from the fact that an increase of energy in the dot corresponds to a decrease in lead $\ell$.~\cite{schaller_single-electron_2013,agarwalla_full_2015} 

We note that more sophisticated QME approaches exist which take into account, \emph{e.g.}, higher-order tunneling processes.~\cite{kaasbjerg_full_2015}
However, Markovian master equations generally fail to address the tunneling-induced level broadening of dot states, and the accurate evaluation of the deeply non-Markovian memory kernels that characterize correlated regimes requires a numerical method on par with iQMC.~\cite{cohen_memory_2011,cohen_numerically_2013,hartle_transport_2015}


\section{Cumulants and thermoelectric quantities}\label{cumulants}

The cumulants of the statistical distribution for particle and energy transfer at the interface of the dot with lead  $\ell$, as well as their cross-correlations, may be obtained from derivatives of the cumulant generating function,
 \begin{equation}
 \mathcal{S}_\ell\left(t;\lambda,\chi\right)=\log{Z\left(t;\lambda,\chi\right)},
 \end{equation}
 with respect to the counting fields $\lambda$ and $\chi$ and with the convention introduced above that these fields are nonzero only for properties stemming from lead $\ell$. In general, we define
 \begin{equation}
C_{\ell,m,k}\left(t\right) = \left.\frac{\partial^{m+k}\mathcal{S}_{\ell}\left(t;\lambda,\chi\right)}{\partial\left(i\lambda\right)^{m}
\partial\left(i\chi\right)^{k}}\right|_{\lambda,\chi=0}.
\label{eq:lambdachi_deriv_def}
\end{equation}
In terms of these derivatives, we may define the $n$-th order cumulants of the separate particle and energy transfer processes
\begin{equation}
\begin{aligned}
C_{\ell,n}^{p}\left(t\right) \equiv C_{\ell,n,0}\left(t\right),
\\
C_{\ell,n}^{E}\left(t\right) \equiv C_{\ell,0,n}\left(t\right).
\label{eq:cumulants_deriv_def}
\end{aligned}
\end{equation}
We may also consider the cross-correlations measuring the statistical influence of particle change on a change in the energy:
\begin{equation}
C_{\ell}^{\left(\times\right)}\left(t\right) \equiv C_{\ell,1,1}\left(t\right).
\label{eq:cross_corr_def}
\end{equation}
One may then evaluate the first and second cumulants of transferred heat from particle, energy and cross-correlation statistics~\cite{kilgour_path-integral_2019} 
\begin{equation}
\begin{aligned}
C_{\ell,1}^{h}\left(t\right) &= C_{\ell,1}^{E}\left(t\right)-\mu_{\ell}C_{\ell,1}^{p}\left(t\right), \\
C_{\ell,2}^{h}\left(t\right) &= C_{\ell,2}^{E}\left(t\right) +\mu_{\ell}^2C_{\ell,2}^{p}\left(t\right) - 2\mu_{\ell}C_{\ell}^{\left(\times\right)}\left(t\right).
\label{eq:heat_cumulants}
\end{aligned}
\end{equation}

In the long time regime, the first and second cumulants yield fluxes, $I_\ell^{p/E/h}$, and zero-frequency 
quantum noises, $S_{\ell}^{p/E/h}$, for particle, energy and heat transfer:~\cite{esposito_efficiency_2015}
\begin{equation}
\begin{aligned}
I_{\ell}^{p/E/h} = \underset{t\rightarrow\infty}{\lim}\frac{C_{\ell,1}^{p/E/h}\left(t\right)}{t},
\\
S_{\ell}^{p/E/h} =\underset{t\rightarrow\infty}{\lim}\frac{C_{\ell,2}^{p/E/h}\left(t\right)}{t}.
\label{eq:steady_state_observables}
\end{aligned}
\end{equation}
Given the quantities above, the overall steady state entropy production rate can be evaluated from the expression~\cite{deffner_nonequilibrium_2011, whitney_thermodynamic_2013, pietzonka_universal_2018} 
\begin{equation}
\sigma=\underset{\ell}{\sum}\beta_{\ell}I_{\ell}^{h}.
\label{eq:entropy_production}
\end{equation}
We note that this relation is only valid for the steady state,~\cite{esposito_nature_2015} and the second law of thermodynamics necessarily implies that $\sigma>0$.


\section{Results}\label{results}
We consider a biased junction, $\mu_L>\mu_R$, both with ($T_L<T_R$) and without ($T_L=T_R$) a temperature gradient. 
In the former case, at appropriately tuned parameters, the system may embody either a heat engine (electron flow against the voltage gradient is driven by the temperature bias) or a heat pump/refrigerator (heat flow against the thermal gradient is driven by the voltage bias).

\subsection{$U=0$ benchmarks and failure of master equations}
We begin with benchmarks of the iQMC method in the noninteracting limit $U=0$ against exact path-integral nonequilibrium Green’s function (PINEGF) results.~\cite{tang_full-counting_2014,yu_full-counting_2016} The iQMC method is numerically exact and one might trivially expect to simply find agreement. Nevertheless, such benchmarks are important whenever a numerical method is generalized, and the iQMC method has not previously been applied to the calculation of energy counting statistics. We furthermore examine the applicability of the QME approach to FCS in the noninteracting limit.

\begin{figure}
\includegraphics{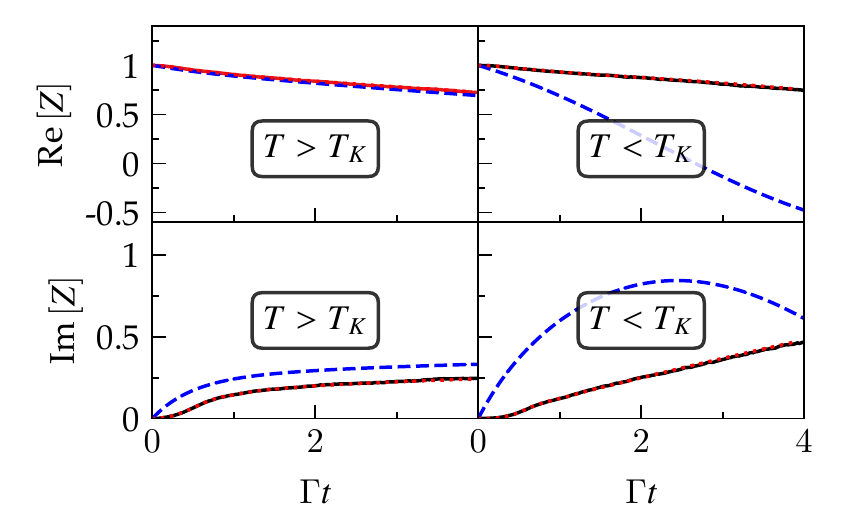}
\caption{Real and imaginary parts of the moment generating function $Z\left(t;\lambda=0.5,\chi=0.1\right)$ in lead $L$ are shown for calculations using the QMC (thick black) QME (dashed blue) and PINEGF (dotted red) methods. We choose the voltage $V=1\Gamma$, interaction strength $U=0\Gamma$ and the empty initial condition. The two leads are held at equal inverse temperature $\beta_{L} = \beta_{R}$. The left panels show the high temperature case $\beta_{L} = 0.4/\Gamma$.
The right panels show the low temperature case $\beta_{L} = 50/\Gamma$. }
\label{fig:QME_vs_TL_TR1.0}
\end{figure}

Fig.~\ref{fig:QME_vs_TL_TR1.0} displays the real (top panels) and imaginary (bottom panels) components of the moment generating function $Z\left(t;\lambda,\chi\right)$ on lead $L$, with counting fields set to $\lambda=0.5$ and $\chi=0.1$.
We note that typically one is most interested in the derivatives of the generating function at the $\lambda,\chi\rightarrow0$ limit, where they give the exact cumulants.
However, logarithmic derivatives of $Z\left(t;\lambda,\chi\right)$ vary slowly with the counting fields, and the finite values of $\lambda$  and $\chi$ chosen here are therefore physically relevant and easier to work with numerically.~\cite{ridley_numerically_2018}
The bias voltage is $V=1\Gamma$ and the initial condition is the empty state $\left|0\right\rangle$.
There is no thermal bias, \emph{i.e.} $T_{L}/T_{R}=1.0$. The left panels are at a higher temperature $\beta_{L}=\beta_{R}=0.4/\Gamma$, whereas the right panels are at a lower temperature $\beta_{L}=\beta_{R}=50/\Gamma$.
Each panel directly compares results from iQMC (black) with QME data (dashed blue) and exact PINEGF data (dotted red).

To within numerical errors, the iQMC and PINEGF methods agree at all times and both temperatures. Notably, the noninteracting limit is a nontrivial benchmark for the iQMC algorithm employed here, which is based on an expansion in the dot--lead hybridization rather than the interaction. As might be expected, the QME fails to correctly capture the time evolution in both cases, and fails qualitatively even at describing the steady state (constant time derivatives at long times) at the lower temperature.~\cite{braggio_full_2006}

\begin{figure}
\includegraphics[width=8.6cm]{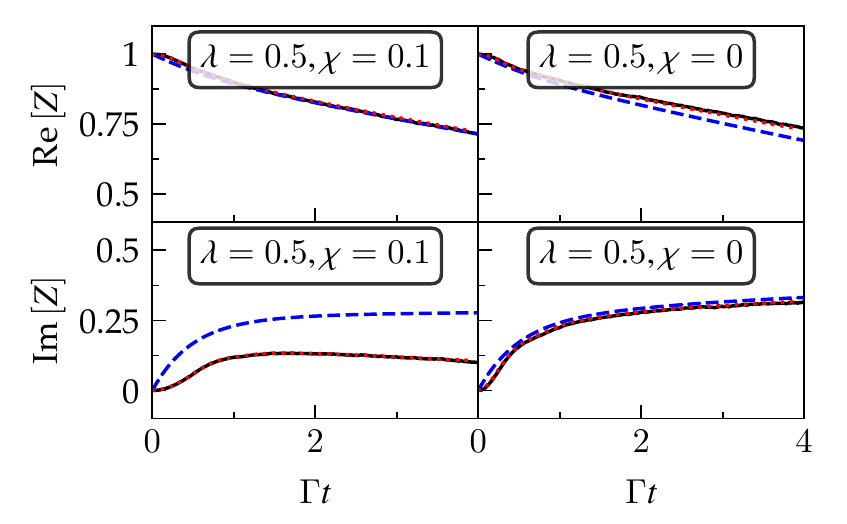}
\caption{Real (upper panels) and imaginary (lower panels) parts of the moment generating function $Z\left(t;\lambda,\chi\right)$ in lead $L$ for the same parameters as in the upper panels of Fig.~\ref{fig:QME_vs_TL_TR1.0}, but with applied thermal bias with $\beta_{L} \beta_{R}$ and left-to-right lead temperature ratio $T_{L}/T_{R}=0.1$. The counting fields are set to $\lambda=0.5$ and $\chi=0.1$ (left panels), and applied temperature bias and $\lambda=0.5$ and $\chi=0$ (right panels).}
\label{fig:QME_vs_TL_TR0.1}
\end{figure}

In Fig.~\ref{fig:QME_vs_TL_TR0.1}, we focus on the higher temperature regime in which the QME is expected to work well, at $\beta_{L}=0.4/\Gamma$.
The left panels repeat the calculation in the left panels of Fig.~\ref{fig:QME_vs_TL_TR1.0}, but with an applied temperature bias $T_{L}/T_{R}=0.1$, so that heat is driven against the direction of the particle current.
Note that this is set up so that the average temperature is higher than $\beta_{L}$, a supposedly even more favorable condition for QME.
Interestingly, however, QME fails to capture the high temperature steady state not just quantitatively, but even in terms of the sign of the long-time derivative (slope) of $\Im{Z}$ (see lower left panel of Fig.~\ref{fig:QME_vs_TL_TR0.1}).

The right-hand panels of Fig.~\ref{fig:QME_vs_TL_TR0.1} show that in this temperature regime the three methods are in excellent agreement for the case of zero energy counting field, $\chi=0$.
We have verified that this remains true in the thermally unbiased case.
Mixed particle--energy counting therefore appears to be more challenging to capture within the QME framework than particle counting alone.

\subsection{Finite-$U$ FCS in QME and iQMC}
Having established that the iQMC method provides reliable FCS in the presence of both the particle and energy counting fields, we continue to explore the interacting case, where no analytical results are available. We concentrate on energy and particle cumulants up to second order, in addition to energy--particle cross correlations $C_{1/2,\ell}^{p/E}\left(t\right)$ and $C_{pE,\ell}^{\left(\times\right)}\left(t\right)$. Using the symmetry properties $Z\left(t;\lambda,\chi\right)=Z\left(t;-\lambda,\chi\right)^{*}$ and $Z\left(t;\lambda,\chi\right)=Z\left(t;\lambda,-\chi\right)^{*}$, all these quantities can be extracted from finite-difference logarithmic derivatives based on just four simulations of the generating function in the $\left(\lambda,\chi\right)$ plane.

\begin{figure}
\includegraphics{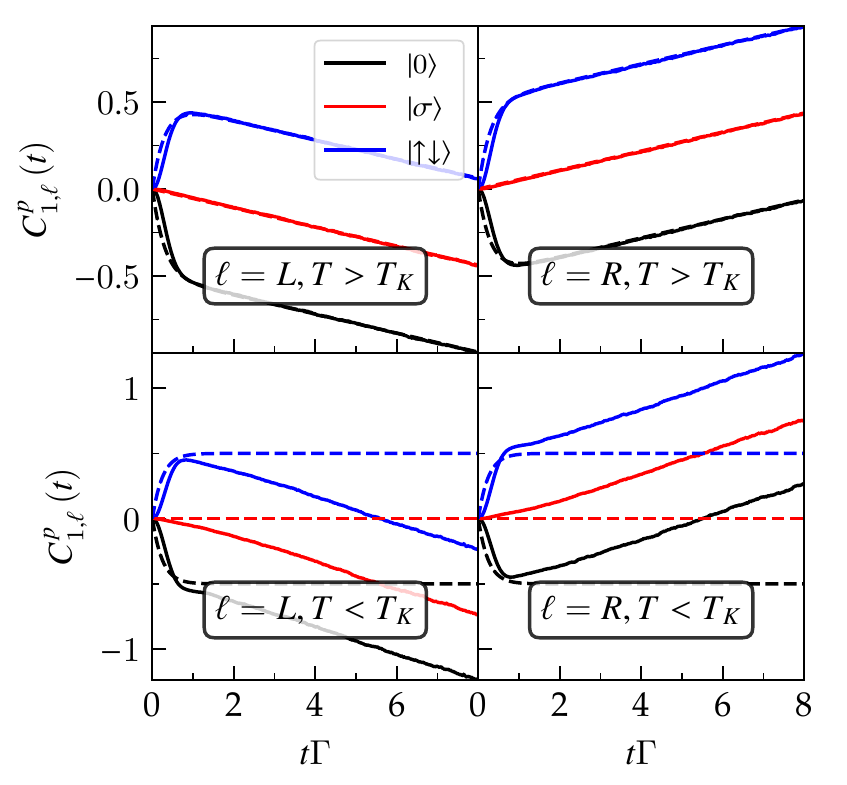}
\caption{First cumulant of particle number change in lead $L$ (left panels) and $R$ (right panels) for different initial conditions, at voltage $V=1\Gamma$, on-site interaction strength $U=8\Gamma$ and with no temperature gradient, $\beta_{L} = \beta_{R}$. The upper panels show the high temperature case of $\beta =  0.4/\Gamma$. The lower panels show the low temperature case of $\beta =  50.0/\Gamma$. Solid lines denote iQMC results, and dashed lines denote QME data.
}
\label{fig:1_Q}
\end{figure}

\begin{figure}
\includegraphics{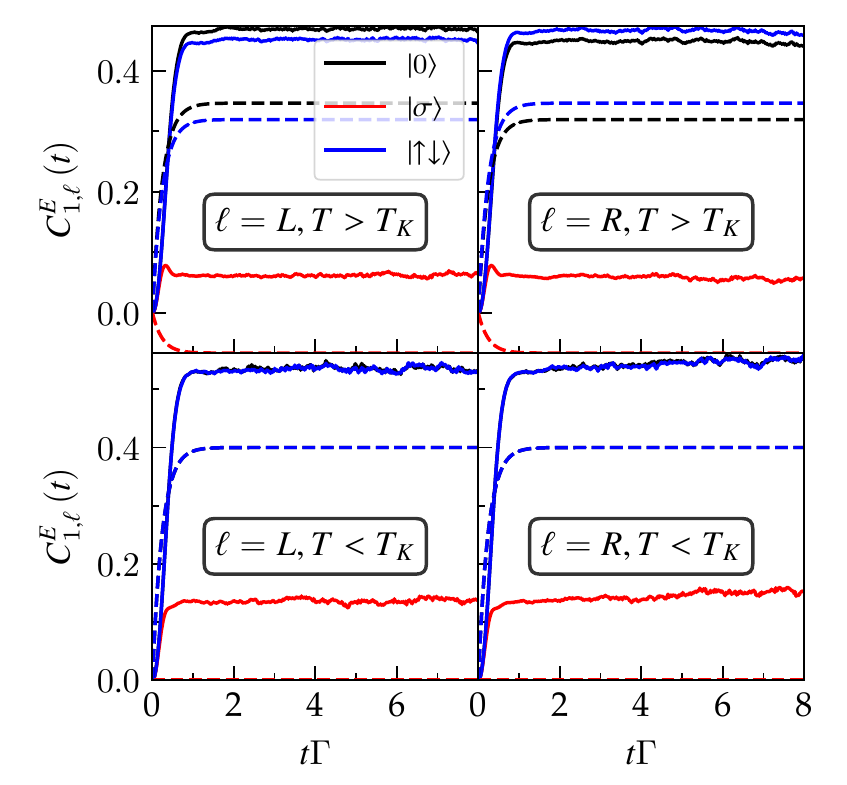}
\caption{
First cumulant of energy change in lead $L$ (left panels) and $R$ (right panels) for different initial conditions, at the same parameters as Fig.~\ref{fig:1_Q}. Solid lines denote iQMC results, and dashed lines QME.
}
\label{fig:1_E}
\end{figure}

\begin{figure}
\includegraphics{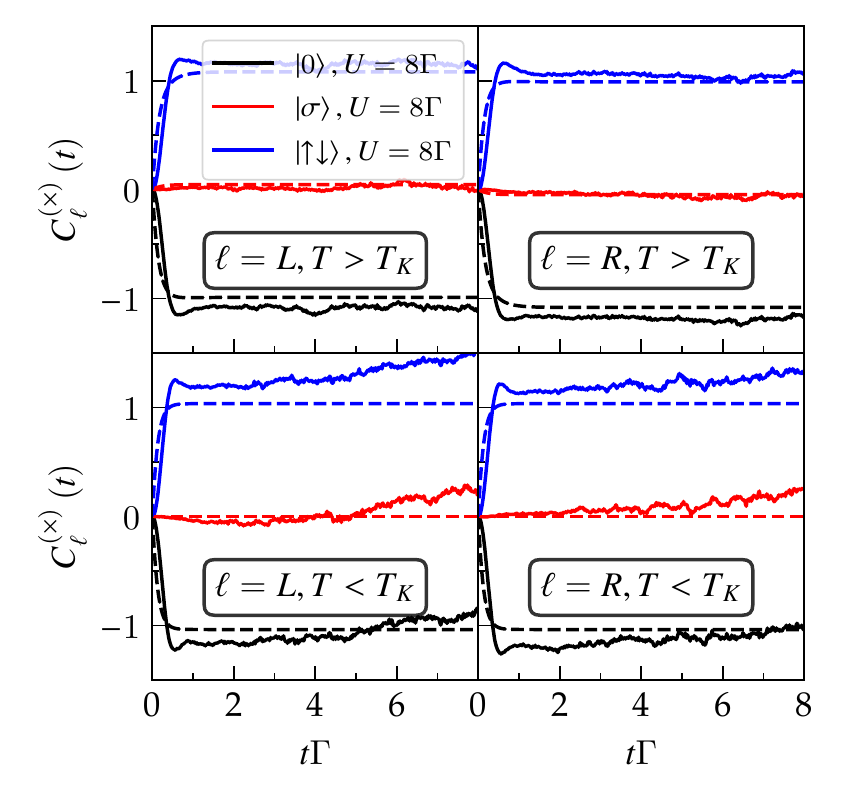}
\caption{
Cross correlations between particle and energy change in lead $L$ (left panels) and $R$ (right panels) for different initial conditions, at the same parameters as Fig.~\ref{fig:1_Q} Solid lines denote iQMC results, and dashed lines QME.
}
\label{fig:pxE}
\end{figure}

In the rest of this section, dashed lines in the figures will denote QME data and solid lines or symbols denote iQMC results. In Figs.~\ref{fig:1_Q}--\ref{fig:C_2_h_compare}, three different initial states of the dot are shown as color-coded curves: empty (black), half-filled (red) and fully-occupied (blue).

Figs. \ref{fig:1_Q} and \ref{fig:1_E}, respectively, show the first cumulants of particle and energy transfer on lead $\ell=L$ (left panels) and $\ell=R$ (right panels), where the counting fields are introduced identically for each lead.

We set the interaction strength to $U=8\Gamma$, the bias voltage to $V=1\Gamma$ and the temperatures to be equal on both leads, such that currents are driven only by the voltage. The approximate Kondo temperature of the system can be estimated as $k_{B}T_{K}\approx0.1\Gamma$, in accordance with the Bethe ansatz formula $k_{B}T_{K}=\sqrt{U\Gamma/2}\exp\left( -\pi U/8\Gamma+\pi\Gamma/2U\right)$.~\cite{hewson_kondo_1993}
We note in passing that recent work proposes nonequilibrium logarithmic corrections to this formula,~\cite{li_corrected_2017} but these corrections are small enough that we are able to safely find temperatures above and below $T_K$ in what follows.
In particular, the upper panels of Figs. \ref{fig:1_Q}--\ref{fig:1_E} are at a high temperature $T=2.5\Gamma \approx 25T_K$, whereas the lower panels are at a low temperature $T=0.02\Gamma \approx T_K / 5$ is shown.

Fig.~\ref{fig:1_Q} shows the total change in electron number with increasing time, from the quench time up until $t=8/\Gamma$. The upper panels shows a remarkable agreement between the iQMC and QME data at high temperatures, with the exception of the short-time regime in which higher-order scattering processes are dominant.~\cite{antipov_voltage_2016}
This plot shows significant initial condition dependence in the transferred charges, which may be understood in terms of Coulomb blockade physics: (i) in the initially unoccupied case (black line), one electron can tunnel onto the dot from the left lead, whose Fermi level is at $\mu_{L}=\Gamma/2$, before the Coulomb repulsion slows down the rate of charge transfer out of this lead, explaining the change to a less negative gradient in $C_{1,L}^{p}\left(t\right)$ after a relaxation time of approximately $\tau_{r}\approx1/\Gamma$.
(ii) When the dot is initially fully occupied (blue line), the particle number may only change on the left lead on the timescale of $\tau_{r}$ by tunneling up the voltage gradient to put the dot in a magnetized state, at which point the Coulomb blockade forces the current to take the same value as in case (i).
This tunneling process is activated by the high temperature $k_{B}T>eV$ of the junction.
(iii) When the dot is initially in the spin-up or spin-down state (red line), the steady state population is immediately established at the quench time, explaining why the gradient of this cumulant is identical to the gradient which only forms after time $\tau_{r}$ has elapsed in cases (i) and (ii).

In the lower panels of Fig.~\ref{fig:1_Q}, the first cumulants are shown for the low temperature ($\beta=50/\Gamma$) case with an otherwise unchanged set of parameters. Here we see a clear temperature-dependent transition in the QME data, as all three dashed lines relax to a constant (zero current) value in a clear signature of Coulomb blockade preventing transport through the dot. However, the iQMC data shows an \textit{enhanced} current magnitude. This is indicative of the additional transport channels enabled by the formation of a Kondo resonance in the nonequilibrium conductance, and illustrates the breakdown of our naive QME approach at low temperatures.

We now turn to the first cumulant of transferred energy for the same system parameters, which is displayed in Fig.~\ref{fig:1_E}.
The QMC data clearly shows that the energy of the left lead undergoes an increase for all initial conditions.
This energy gain is uniform across the entire junction, as can be seen by considering the cumulants in the right lead, shown in the right-hand panels of Fig.~\ref{fig:1_E}.
It is therefore not associated with the direction of flow of particle current, which has a different sign in each lead.
Instead, this energy increase can be interpreted as a charging effect, \textit{i.e.} it arises from the change in the electrostatic energy driven by the addition or removal of an electron from the dot.

We point out that our data on energy cumulants (and later cross correlations) is significantly noisier than the corresponding results for particle cumulants.
The main reason is that in order to obtain converged results for these cumulants from the logarithmic finite difference derivatives, relatively small values of $\chi$ were needed.
Since the absolute value of the differences from the $\chi=0$ values is therefore small, this results in large relative errors when evaluating numerical logarithmic derivatives with respect to $\chi$.
We do not include rigorous uncertainty estimates, which are expensive to obtain in this case.\cite{cohen_taming_2015,antipov_currents_2017}
As a rough guideline, by comparing uncorrelated runs (not shown) we find that the data is reliable to within a small factor (say, ~2--3) of the apparent stochastic noise.

Whereas the total transferred energy at low temperature is equal for the initially fully occupied and empty states, when the temperature is raised this symmetry is broken.
This can be seen in the gap that opens up between the blue and black lines in Fig.~\ref{fig:1_E} in both the iQMC and the QME data.
Also for high temperatures, the symmetry between the energies transferred to the left and right leads is broken, such that the order of the blue and black lines is reversed between $L$ and $R$.
To understand this, it is enough to consider the rates for dot--lead transfer processes in the $L$- and $R$-lead cases separately.
In order to magnetize the dot when the initial condition is $\left|0\right\rangle$ or $\left|\uparrow\downarrow\right\rangle$, an energy of $-4\Gamma$ must be transferred to the dot from one of the lead states $k$.
In the initially empty state, a particle is transferred out of one lead $\ell$ and the corresponding transfer rate $R_{\ell}$ is proportional to the Fermi function $f\left(-U/2-\mu_{\ell} \right)$.
In the initially fully occupied state the rate is proportional to the hole occupation $f\left(U/2-\mu_{\ell} \right)$.
At infinite $k_{B}T$, $R_{L}=R_{R}=1/2$ for both initial conditions, so there is no symmetry breaking in the system.
In the opposite regime of $k_{B}T=0$, $R_{L}=R_{R}=1$ for the empty state and $R_{L}=R_{R}=0$ for the full state, once again symmetric.
However, at \emph{intermediate} temperatures $k_{B}T\sim\left|\triangle E\pm V\right|$, the left--right symmetry is broken.
In this case, $R_{L}>R_{R}$ for the empty state and $R_{L}<R_{R}$ for the full state.
If the system is initially empty, energy transfer to the left lead is favored in the short time regime.
The converse is true for the full initial state.

The QME correctly captures the qualitative, though not the quantitative, aspects of the high and intermediate temperature energy transfer due to charge dynamics.
Notably, however, it fails to predict even the direction of overall energy transfer in the initially magnetized case.
This can be seen by comparing the dashed and solid red curves in the top panel of Fig.~\ref{fig:1_E}.
Less surprisingly, QME also spuriously predicts no energy transfer in this case at low temperatures.
This is because at the level of QME, the system essentially begins in its steady state under such conditions.

In Fig.~\ref{fig:pxE} we present the time evolution of the statistical correlation between the changes in particle number and energy in each lead, $C_{\ell}^{\left(\times\right)}\left(t\right)=\left\langle \triangle \epsilon_{\ell}\left(t\right)\triangle n_{\ell}\left(t\right)\right\rangle$.
Immediately one observes that the sign of this quantity depends on the initial condition, making it an excellent observable for experimental detection of the initial system preparation.
In the case of an initially occupied (unoccupied) dot, the particle number in the leads increases (decreases) during the transient regime $t< \tau_{r}$.
However, for either initial condition, the total energy in the leads undergoes the same increase, as shown in Fig.~\ref{fig:1_E}.
This results in the dependence of sign on initial condition in Fig.~\ref{fig:pxE}.
At long times and low temperature (bottom panels), $C_{\ell}^{\left(\times\right)}\left(t\right)$ increases, hinting at a buildup of correlations between charge and energy transport that is completely absent within the QME.

\subsection{Heat transfer statistics and thermodynamics}

\begin{figure}
\includegraphics{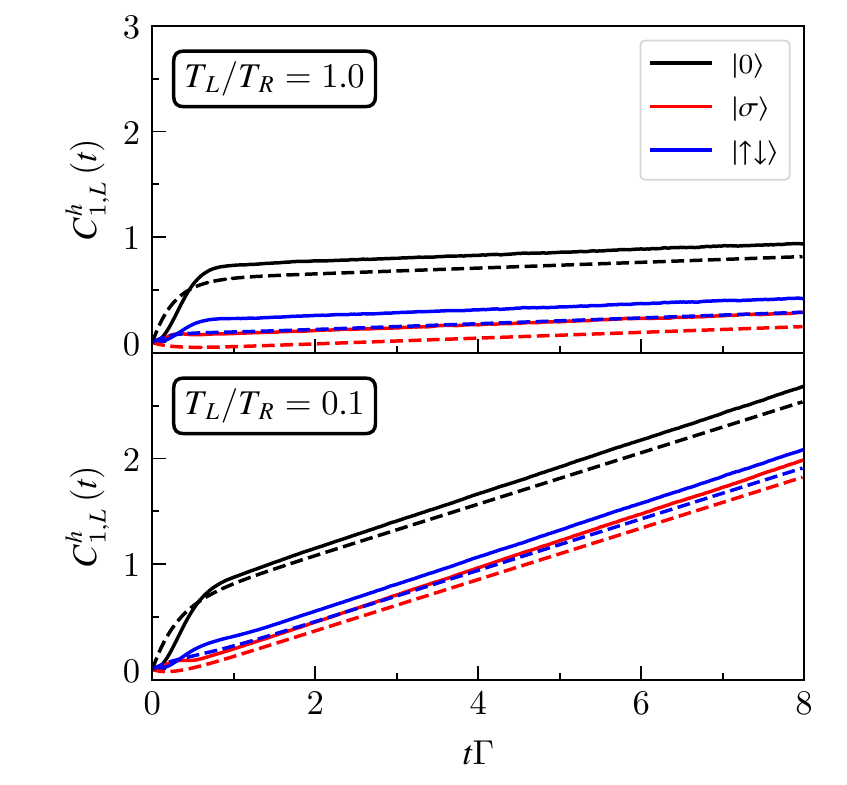}
\caption{First cumulant of heat transfer into lead $L$ for different initial conditions, at voltage $V=1\Gamma$ and interaction strength $U=8\Gamma$, and for the iQMC (thick lines) and QME (dashed lines) methods. We show the high temperature regime of $\beta_{L} = 0.4/\Gamma$, with the temperature ratios set to $T_{L}/T_{R}=1.0$ (upper panel) and $T_{L}/T_{R}=0.1$ (lower panel). Solid lines denote iQMC results, and dashed lines QME.}
\label{fig:C_1_h_compare}
\end{figure}

We now examine the thermodynamic effects of simultaneously applying both a voltage bias and a temperature gradient.
We will first focus on the flow of heat (see Eq.~\eqref{eq:heat_cumulants}) and its fluctuations through the system.
In Fig.~\ref{fig:C_1_h_compare} the first cumulant of heat transfer, $C_{1,L}^{h}\left(t\right)$, is shown in the presence of a voltage bias of $V=1\Gamma$.
The left lead is held at a high temperature $T_{L}=2.5\Gamma$, while $T_R$ is set so as to enforce a temperature ratio of either $T_{L}/T_{R}=1.0$ (upper panel) or $T_{L}/T_{R}=0.1$ (lower panel).
Therefore, in the second case, a thermal bias is applied in a direction opposite to that of the voltage bias.
We observe that while the left lead is heated either with or without a temperature bias, this heating is greatly enhanced when energy is driven into it from the hotter right lead.
In addition, mild deviations between the iQMC and QME persist for the the gradients of the cumulants even at long times, even though both leads are at a high temperature.
This occurs for the same reasons pointed out in the discussion of Fig.~\ref{fig:QME_vs_TL_TR0.1}.

\begin{figure}
\includegraphics{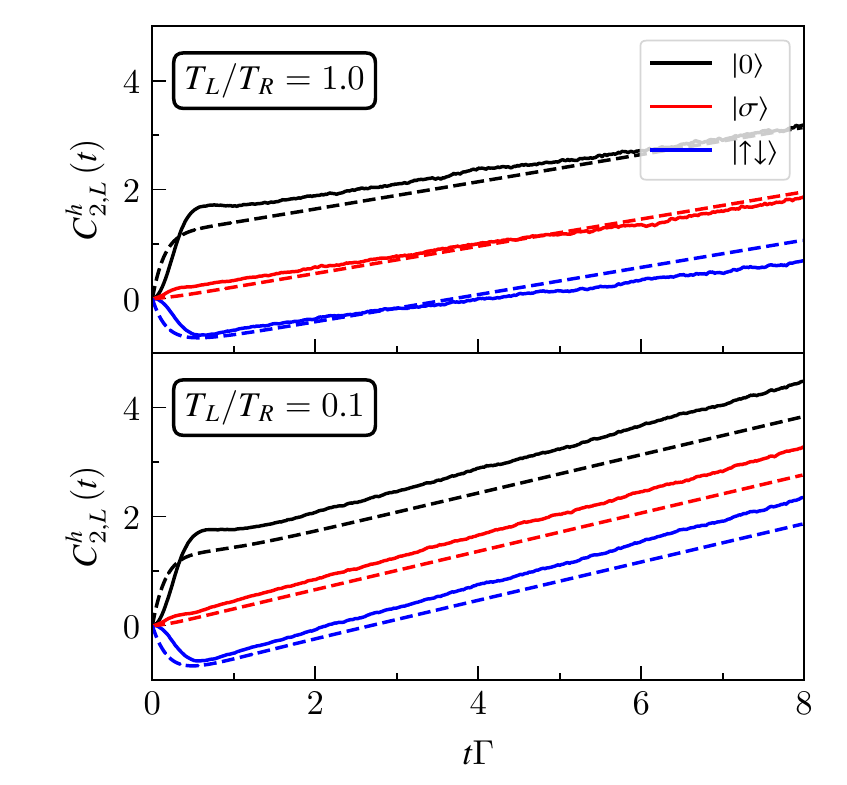}
\caption{Second cumulants of heat transfer for the same parameters as in Fig.~\ref{fig:C_1_h_compare}.  Solid lines denote iQMC results, and dashed lines QME.}
\label{fig:C_2_h_compare}
\end{figure}

In Fig.~\ref{fig:C_2_h_compare} the second cumulants of heat transfer are shown for the same parameters as in Fig.~\ref{fig:C_1_h_compare}.
The thermal fluctuations are rather large even without a temperature bias, and their enhancement by this bias is not as strong.
The QME is not able to capture these fluctuations as well as it captures the mean heat transfer, resulting in more significant deviations from the numerically exact data.
These deviations are more obvious in the presence of a thermal bias, despite the larger average temperature of the system.

\begin{figure}
\includegraphics{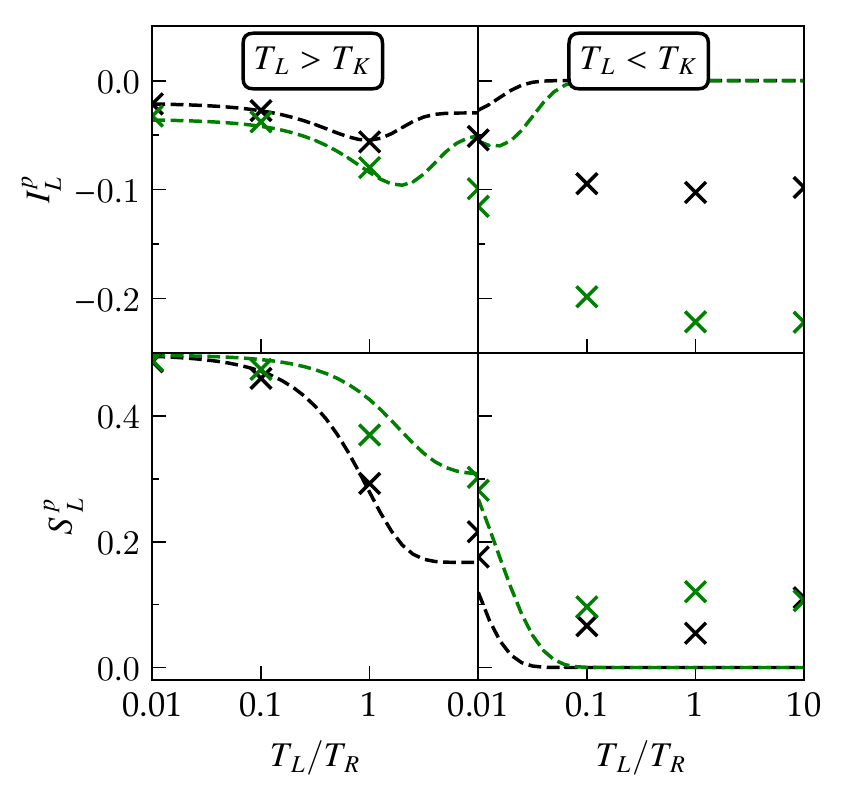}
\caption{Steady steady particle current (upper panels) and current noise (lower panels) as a function of temperature ratio $T_{L}/T_{R}$. QME data is shown at $U=8\Gamma$ (dashed black) and $U=4\Gamma$ (dashed green). Black crosses indicate iQMC data at $U=8\Gamma$, and green crosses denote the iQMC results for $U=4\Gamma$. The left lead is held at constant high (left panels) or low (right panels) temperature, and the bias voltage is fixed at $V=1\Gamma$.}
\label{fig:current_noise_Tcompare}
\end{figure}

In our analysis so far, we have considered the entire time evolution following the coupling quench.
We now investigate the steady state thermodynamics that eventually emerges.
This is extracted from the long-time dynamics of the cumulants (see Eq.~\eqref{eq:steady_state_observables}).
In Fig.~\ref{fig:current_noise_Tcompare} we present the particle current (upper panels) and the associated shot noise (lower panels), plotted on a logarithmic scale as functions of the left--right lead temperature ratio $T_{L}/T_{R}$ at constant $T_L$.
$T_L$ takes on the same two values that were considered in previous plots, one of which (shown in left panels) is above the Kondo temperature at $U=8\Gamma$ and the other of which (right panels) is below it.
The right lead temperature varies from $100T_{L}$ to $0.1T_{L}$.
QME data is shown at interaction strength $U=8\Gamma$ (black dashed line) and $U=4\Gamma$ (green dashed line), and iQMC data at $U=8\Gamma$ (black crosses) and $U=4\Gamma$ (green crosses).
Our estimate for the magnitude of numerical uncertainties~\cite{cohen_taming_2015,antipov_currents_2017} is smaller than the symbol size.

The definition we have used for the particle current can be interpreted as the flow rate into the left lead.
Its sign at the parameters considered here is always negative, \textit{i.e.} particles are flowing from left to right, consistent with the directionality induced by the voltage gradient.
Within the QME picture, the system is in the  Coulomb blockade regime, such that reducing the interaction strength consistently increases the magnitude of the particle current and noise (\textit{cf.} green and black dashed curves in Fig.~\ref{fig:current_noise_Tcompare}).

As might be expected, the QME successfully reproduces the numerically exact iQMC results for both current and noise at high (left and right) temperatures.
It therefore proves to be an excellent approximation in the left part of the left panels (high $T_L$ and $T_R$), and a reasonable one at the right edge of these panels (high $T_L$ and low $T_R$).
In the low temperature case where $\beta_{L}=50/\Gamma$, the QME fails qualitatively.
In particular, the QME predicts complete suppression of both current and noise when both leads are at low temperatures.
In the physical regime explored here, as the temperature decreases, this may first be attributed to higher-order scattering processes, and later to the formation of a correlated Kondo transport channel.
Nonmonotonic behavior can be observed in the iQMC results at high $T_L/T_R$.
Here, the equilibrium system might be expected to be driven deeper into the Kondo regime by the lower average temperature, but this effect competes with the nonequilibrium fluxes that eventually break down the Kondo singlet.

\begin{figure}
\includegraphics{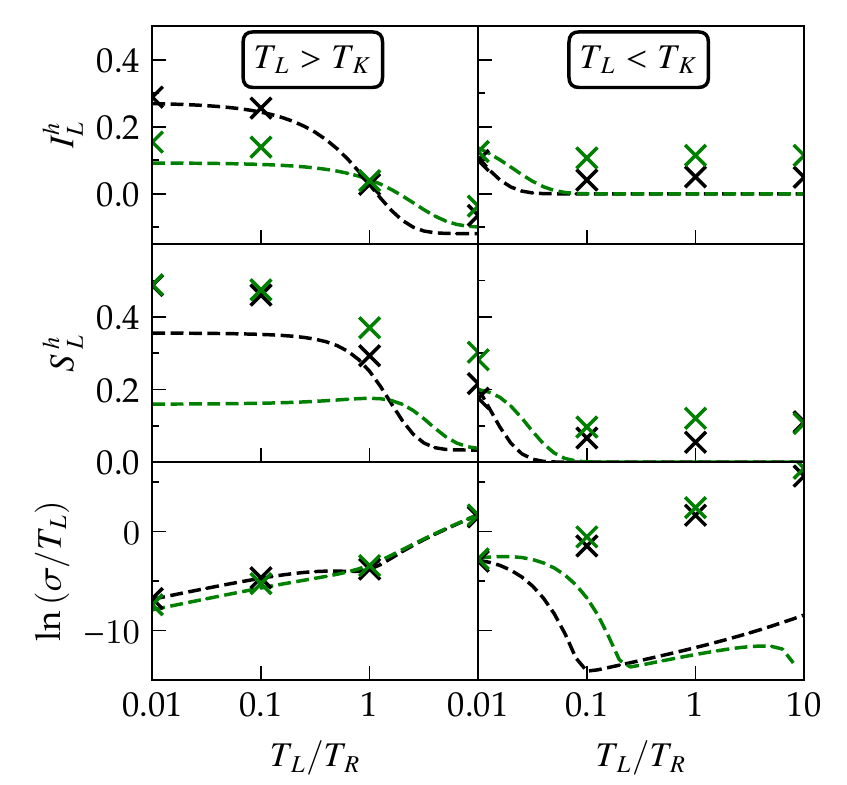}
\caption{Steady state heat current (upper panels), heat current noise (middle panels) and natural logarithm of the entropy production rate normalized by the left lead temperature (lower panels), plotted as functions of the temperature ratio $T_{L}/T_{R}$. All parameters and labelling conventions are as in Fig.~\ref{fig:current_noise_Tcompare}.}
\label{fig:heat_entropy_Tcompare}
\end{figure}

In Fig.~\ref{fig:heat_entropy_Tcompare}, we move on to examine the steady state values of heat current (upper panels), heat current fluctuations (middle panels) and the logarithm of the entropy production rate normalized by the left lead temperature (lower panels). In these plots the entropy is temperature-normalized so that the $T_{L}>T_{K}$ and $T_{L}<T_{K}$ plots may be compared on the same scale, and we consider the logarithm because the growth of entropy production with increasing average temperature is known to be exponential.
As before, we consider temperatures $T_L=2.5\Gamma>T_K$ (left panels) and $T_L=0.02\Gamma<T_K$ (right panels).

At high temperature (left panels), QME provides a good estimate of the heat current, as it did for the particle current.
QME consistently underestimates the fluctuations in heat current, but does predict the correct trend in that fluctuations decrease when the right lead is cooled down.
However, it also predicts a strong and spurious dependence on $U$ at low temperature ratios, which is not observed in the iQMC results.
A reversal in the direction of the steady state heat current as a function of the temperature ratio, as illustrated in Fig.~\ref{fig:nanojunction}, is observed in both methods (upper left panel).
The temperature ratio at which this occurs is indistinguishable from 1 within our numerical resolution, indicating that the direction of heat flux is completely determined by that of the temperature gradient in this parameter regime.

The QME picture for the high temperature entropy generation $\sigma$ (lower panel of Fig.~\ref{fig:heat_entropy_Tcompare}) is surprisingly accurate, though this is partially due to the logarithmic scale.
A monotonic and approximately exponential growth with the temperature ratio is observed, with different exponents on the two sides of $T_L/T_R=1$.

For low $T_L$ (right panels of Fig.~\ref{fig:heat_entropy_Tcompare}, QME fails qualitatively except where $T_R$ is large.
The change in sign at $T_L/T_R=1$ observed at high temperature vanishes at low temperature.
In the QME approximation, this occurs simply because the heat current vanishes entirely.
The iQMC result does show a finite heat current which remains positive for $T_R<T_L$, such that the left lead continues to heat in this case.
This effect and the wrong QME result can be understood by considering the two components of the heat current (see Eq.~\eqref{eq:heat_cumulants}) separately.
At such low lead temperatures, the energy current (not shown) is essentially zero, a fact that is captured by QME; while the particle current continues to be carried by the Kondo channel, not captured by QME.
The lack of sign reversal in the heat current is accompanied by an increase in its fluctuations, again in opposition to the high temperature result.
This may be interpreted as a precursor to an eventual reversal that may manifest if the right lead is cooled further, or if the voltage bias is reduced.

Unlike at high temperatures, the QME estimate for the low temperature entropy production (lower right panel of Fig.~\ref{fig:heat_entropy_Tcompare}) fails catastrophically.
Here, it predicts a spurious structure with multiple peaks and troughs instead of the correct behavior, which is essentially exponential.
At the lower temperature, the difference in exponents at the two sides of $T_L/T_R=1$ is either nonexistent or too small to detect within our numerical resolution.
The detectable exponent is clearly larger than that appearing in the high temperature case, as is the overall entropy production rate.
This result is most likely connected to the opening of the Kondo transport channel, and invites further theoretical analysis beyond the scope of the present work.


\section{Conclusions}\label{conclusion}

We presented a numerically exact method for evaluating full counting statistics (FCS) in nonequilibrium quantum junctions based on the inchworm quantum Monte Carlo (iQMC) approach.
The method accounts for both particle and energy transport statistics, and is applicable in a wide range of parameters that includes the strongly correlated Kondo regime.

We benchmarked the method against the nonequilibrium Green's function formalism in the noninteracting case.
We also carried out an extensive comparison between the iQMC and a simple quantum master equation (QME) approach at different temperature regimes, showing clearly where the latter method fails.
Surprisingly, we found that the QME approximation fails to produce the correct energy counting statistics even at temperatures significantly higher than $T_{K}$, although agreement between QME and iQMC improves as the temperature tends to infinity.
We further found that the presence of a thermal gradient across the molecular junction makes the agreement worse even at high temperatures, and when the gradient increases the overall average temperature.

At temperatures above the Kondo threshold, we found that there is a left--right symmetry breaking in the system that can be observed in the first cumulant of energy transfer: energy is distributed among the leads in away that depends on the initial preparation of the system.
When the temperature is lowered below $T_{K}$, the iQMC calculations predict finite values for the particle and heat current and noise, whereas the QME method predicts full suppression of both current and noise due to the Coulomb blockade effect.
In general, the disagreement between the QME and iQMC approaches is most significant in the noise and the energy--particle cross-correlations, confirming that noise measurements offer more sensitive probes of higher-order scattering processes and many-body correlations than average currents.

Finally, we investigated the steady state entropy production rates at different interaction strengths using both the Monte Carlo and master equation methods.
Among other things, we found that the entropy production rate from master equations spuriously predicts an opposite trend to that computed from iQMC as the average temperature of the system is reduced significantly below $T_{K}$.

This paper provides the basis for future investigations into several more specific questions, including the FCS of energy transport in periodically-driven systems and the properties of levitons in the Kondo regime, as in Ref.~\onlinecite{suzuki_coherent_2017}.
Another interesting direction is models with multiple orbitals, where QME may fail at any temperatures due to an inability to properly account for bath induced coherences in the system.
In general, the tools presented here can to study a variety of fundamental questions in quantum thermodynamics, by obtaining the time-dependence of entropy production, testing thermodynamic uncertainty relations and validating fluctuation--dissipation relations in strongly correlated quantum many-body systems.

\acknowledgments{
G.C. acknowledges support by the Israel Science Foundation (Grant No. 1604/16). E.G. was supported by DOE ER 46932. 
M.G. acknowledges support by the National Science Foundation (Grant CHE-1565939). 
International exchange and collaboration was supported by Grant No. 2016087 from the United States-Israel Binational Science Foundation (BSF) and by University of California San Diego.}

\bibliographystyle{apsrev4-1}
%


\end{document}